\documentclass[%
 reprint,
 amsmath,amssymb,
 aps,
prb,
]{revtex4-2}

\usepackage{graphicx}
\usepackage{dcolumn}
\usepackage{bm}
\usepackage{braket}
\usepackage[colorinlistoftodos]{todonotes}
\usepackage[T1]{fontenc}
\usepackage{url}
\usepackage{hyperref}
\usepackage{amssymb}
\usepackage{hyperref}
\usepackage{multirow}

\begin{document}


\title{Nonlinear eigenvalue algorithm for \texorpdfstring{$GW$}{Lg} quasiparticle equations}

\author{Dongming Li}
 \email{dongmingli@umass.edu}

\author{Eric Polizzi}
 \email{epolizzi@engin.umass.edu}
 \altaffiliation[Also at ]{Department of Mathematics and Statistics, University of Massachusetts Amherst.}
\affiliation{
 Department of Electrical and Computer Engineering, University of Massachusetts, Amherst, MA, United States.
}
%





\begin{abstract}
The use of Green's function in quantum many-body theory often leads to nonlinear eigenvalue problems, as the Green's function needs to be defined in energy domain. The \texorpdfstring{$GW$}{Lg} approximation method is one of the typical examples. In this article, we introduce a method based on the FEAST eigenvalue algorithm for accurately solving the nonlinear eigenvalue \texorpdfstring{$G_0W_0$}{Lg} quasiparticle equation. Based on the contour integral method for nonlinear eigenvalue problems, the energy (eigenvalue) domain is extended to complex plane. Hypercomplex number is introduced to the calculation of \texorpdfstring{$GW$}{Lg} self-energy to carry imaginary parts of both Green's functions and FEAST contour quadrature nodes. Calculation results for various molecules are presented and compared with the more conventional graphical solution approximation method and spectral functions method. It is confirmed that the Highest Occupied Molecular Orbital (HOMO) from the Kohn-Sham (and Hartree-Fock) equation is very close to that of \texorpdfstring{$GW$}{Lg}, while the Least Unoccupied Molecular Orbital (LUMO) shows noticeable differences.
\end{abstract}


\maketitle

The $GW$ approximation has established itself as a cornerstone in the realm of many-body quantum theory, providing a robust framework for the calculation of quasiparticle (QP) properties in a wide range of materials. Originating from the pioneering work of Lars Hedin in 1965 \cite{Hedin1965}, the $GW$ method has undergone extensive development, culminating in a versatile tool for investigating electronic excitations. The method derives its name from the product of the Green’s function ($G$) and the screened Coulomb interaction ($W$), which collectively account for the exchange-correlation effects beyond the Hartree-Fock approximation.

The $GW$ approximation emerged as a response to the limitations of simpler models like Hartree-Fock and density functional theory (DFT) in describing excited states. Hedin’s formalism provided a systematic way to improve upon these methods by incorporating dynamic screening effects. Initially, practical applications of the $GW$ method were limited by computational constraints. However, advancements in numerical techniques and computational power have facilitated its application to increasingly complex systems.
Early applications of $GW$ were focused on simple semiconductors and insulators \cite{Louie1986,Godby1986}, demonstrating significant improvements over DFT in predicting band gaps and electronic spectra. The advent of iterative and self-consistent $GW$ schemes in the 1990s \cite{Barth1996,Holm1998,Schone1998}, 2000s \cite{Gonzalez2001,Shishkin2006,Stan2009} and 2010s \cite{Rostgaard2010,Blase2011,Caruso2012,Caruso2013,Marom2012,Wilhelm2016,Grumet2018}, notably the ev$GW$ and sc$GW$ approaches, marked a significant leap in accuracy and reliability.

In recent years, the $GW$ approximation has seen substantial advancements in both theoretical formulations and computational implementations. Hybrid methods that combine $GW$ with other techniques, such as $GW$+BSE (Bethe-Salpeter Equation) \cite{Salpeter1951,Strinati1982,Louie2000,Sham1980,Komsa2012,Ramasubramaniam2012,Huser2013,Shi2013,Qiu2016,Wu2015,Korbel2014,Bruneval2015,Jacquemin2015} and $GW$+DMFT (Dynamical Mean-Field Theory) \cite{Georges1992,Biermann2003,Kotliar2006,Biermann2014,Choi2016,Nilsson2017,Zhu2021}, have expanded the applicability of $GW$ to more complex systems, including strongly correlated materials. The development of efficient and high-performance computing  algorithms has further enhanced the feasibility of $GW$ calculations for large-scale systems. \cite{Gao2016,Gao2024,Kim2020,Shao2016,Shao2018}

Despite these advancements, $GW$ implementations often rely on linearized eigenvalue equations and DFT Kohn-Sham (KS) wavefunction approximations, which may limit their accuracy in certain scenarios. It is claimed that the KS wavefunctions are often very close to the $GW$ wavefunctions, but we cannot yet say that it is correct for all cases. Nonlinear eigenvalue algorithm, by contrast, offer a more precise framework for capturing the full extent of QP interactions. The eigenvalue-dependent nonlinear eigenvalue approach directly tackles the energy term inside the self-energy in the QP equations, avoiding approximations that can lead to inaccuracies in QP energies. The formulation of the $GW$ QP equation as a nonlinear eigenvalue problem involves solving the Dyson equation, where the self-energy depends on the QP energies themselves.

In this work, we provide accurate solutions (eigen-paris) to the $G_0W_0$ (one-shot $GW$) QP equation, which is the most widely used scheme and the starting point of the self-consistent $GW$ calculations.
Our approach  leverages recent advancements in the FEAST algorithm \cite{Polizzi2009,Gavin2018,Brenneck2020}
to solve the nonlinear eigenvalue QP equation without any numerical approximations.

We begin by outlining the theoretical framework of the $G_0W_0$ approximation and its formulation as a nonlinear eigenvalue problem.

\section*{non-linear formulation of \texorpdfstring{$G_0W_0$}{Lg}}

The $G_{0}W_{0}$ approximation quasi-particle equation in atomic units is given as
\begin{equation}
  \label{eq1}
    \begin{split}
\hat{h}\psi(r)&+\int^{+\infty}_{-\infty}\Sigma(r,r',\varepsilon)\psi(r')dr'=\varepsilon\psi(r),\\
&\mbox{with}\;\;\;\hat{h}=-\frac{1}{2}\nabla^2+v_{ext}(r)+v_{H}(r),
    \end{split}
\end{equation}
where $v_{ext}$, $v_{H}$ correspond to the external and the Hartree potentials, respectively. The term $\Sigma(r,r',\varepsilon)$ is the exchange-correlation self-energy which contains the many-body quantum effects information. Eq.~(\ref{eq1}) is clearly a non-local and non-linear eigenvalue equation
since the self-energy term contains all of these two difficult-to-handle features. The self-energy in one-shot $GW$ approximation is given by the convolution integral of the Green's function ($G_0$) and the screened Coulomb interaction ($W_0$) in energy domain.
\begin{equation}
  \label{eq2}
    \begin{split}
\Sigma(r,r',\varepsilon)&=\\
\frac{\mathbf{i}}{2\pi}&\int^{+\infty}_{-\infty} e^{\mathbf{i}\omega\eta}G_{0}(r,r',\varepsilon+\omega) W_{0}(r,r',\omega)d\omega.
    \end{split}
\end{equation}
An exponential term $e^{\mathbf{i}\omega\eta}$ with a positive infinitesimal $\eta$ is inserted into the integral to make sure that the integral is convergent in the complex plane and physically meaningful. The $G_{0}$ means mean-field one-body Green's function that could be obtained by a mean-field Hamiltonian, such as DFT, and $W_{0}$ is the screened Coulomb interaction which is determined by the mean-field one-body Green's function $G_{0}$ again at the Random Phase Approximation (RPA) level and given by 
\begin{equation}
W(r,r',\omega)=\int^{+\infty}_{-\infty} \epsilon^{-1}(r,r'',\omega) v(r'',r')dr'',
\end{equation}
where $v(r,r')=1/|r-r'|$ is the bare Coulomb interaction, and the dielectric function $\epsilon(r,r',\omega)$ is calculated as follows: 
\begin{equation}
  \label{eq4}
\epsilon(r,r',\omega)=\delta(r,r')-\int^{+\infty}_{-\infty} v(r,r'')\chi_{0}(r'',r',\omega)dr'',
\end{equation}
the irreducible polarizability, $\chi_{0}$, in Eq.~(\ref{eq4}), is the convolution integral of two mean-field one-body Green's functions usually with a factor 2 to represent the spin - $\sigma$ for the closed shell systems:
\begin{equation}
  \label{eq5}
    \begin{split}
&\chi_{0}(r,r',\omega)=\\
&-\frac{\mathbf{i}}{2\pi}\sum_{\sigma}\int^{\infty}_{-\infty} e^{\mathbf{i}\omega'\eta} G_{0}(r,r',\omega+\omega')G_{0}(r',r,\omega') d\omega'.
    \end{split}
\end{equation}
It can be seen from Eq.~(\ref{eq2}) to (\ref{eq5}) that the self-energy is a energy domain integral that is solely determined by two physical quantities, the bare Coulomb interaction $v$ and the Green's function $G_{0}$. Therefore, 
the construction of the Green's function and the evaluation of the energy integral are essential for $GW$ calculations.
For Green's function, typically, the most commonly used method is the eigenfunction decomposition, another way is to construct the mean-field Hamiltonian -- $\hat{H}_{0}$, compute the resolvent (inverse) directly and add the remainder imaginary term.
\begin{equation}
    \label{eq6}
    \begin{split}
G_0(r,r',\varepsilon)&=\sum_{n}\frac{\psi_{n}(r)\psi_{n}^{*}(r')}{\varepsilon-\varepsilon_{n}-\mathbf{i}\eta\; sgn(E_{F}-\varepsilon_{n})}\\
&=(\varepsilon+\mathbf{i}\eta-\hat{H}_{0})^{-1}+R(\mathbf{i}\eta),
    \end{split}
\end{equation}
where $\varepsilon_{n}$ and $\psi_{n}$ are the eigenvalues and eigenfunctions of $\hat{H}_{0}$, $E_F$ is the Fermi energy level of the system $\hat{H}_{0}$ and $R(\mathbf{i}\eta)$ could be computed using the occupied orbitals only. 

Using eigenfunction decomposition, Eq.~(\ref{eq5}) can also be equivalently rewritten as the Lehmann representation. Alternatively, Eq.~(\ref{eq5}) can be solved analytically and expressed by the Green's function and the occupied orbitals. In this work, by using the second equations in both (\ref{eq6}) and (\ref{eq7}), we eliminated the need to solve for unoccupied orbitals. This allows us to compute the Green's function $G_0$ and the irreducible polarizability $\chi_0$ using only the occupied orbitals.
\begin{equation}
    \label{eq7}
    \begin{split}
\chi_{0}(r,r',\omega)=\sum_{\sigma}\sum_{i}^{occ}&\sum_{a}^{unoc}\Bigl[\frac{\psi^{*}_{i}(r)\psi_{a}(r)\psi_{i}(r')\psi^{*}_{a}(r')}{\omega-(\varepsilon_{a}-\varepsilon_{i})+2\mathbf{i}\eta)}\\
&~~~-\frac{\psi_{i}(r)\psi^{*}_{a}(r)\psi^{*}_{i}(r')\psi_{a}(r')}{\omega+(\varepsilon_{a}-\varepsilon_{i})-2\mathbf{i}\eta)}\Bigr]\\
=\sum_{\sigma}\sum_{i}^{occ}\psi_{i}(r)&\psi_{i}^{*}(r')\Bigl(G_0(r,r',\varepsilon_i
+\omega+\mathbf{i}\eta)\\
&~~~+G_0(r',r,\varepsilon_i-\omega+\mathbf{i}\eta)\Bigr).
    \end{split}
\end{equation}

In practice, exchange and correlation parts of the self-energy are treated separately, 
$\Sigma(\varepsilon)=\Sigma^{X}+\Sigma^{C}(\varepsilon)$
, using the following two expressions for the exchange part $\Sigma^{X}$, and the correlation part $\Sigma^{C}$:
\begin{equation}
    \label{eq8}
  \begin{split}
\Sigma^{X}(r,r')&=\frac{\mathbf{i}}{2\pi}\int G_{0}(r,r',\varepsilon+\omega') v(r,r')d\omega'\\
&=-\sum_{j}^{occ}\frac{\psi_{j}(r)\psi^{*}_{j}(r')}{|r-r'|},
    \end{split}
\end{equation}
\begin{equation}
  \label{eq9}
  \begin{split}
&\Sigma^{C}(r,r',\varepsilon)=\\
\frac{\mathbf{i}}{2\pi}&\int G_{0}(r,r',\varepsilon+\omega) (W_{0}(r,r',\omega)-v(r,r'))d\omega\\
&=\frac{\mathbf{i}}{2\pi}\int G_{0}(r,r',\varepsilon+\omega) W_{0}^{C}(r,r',\omega)d\omega.
    \end{split}
\end{equation}
It could be noted that the exchange part (\ref{eq8}) has an analytical solution and becomes energy independent, leaving only the correlation part (\ref{eq9}) energy dependent (for simplicity, the exponential term and the infinity notations will be omitted from now on).

Several different approaches are available to carefully evaluate the correlation component, including: Analytical Continuation (AC) \cite{Ren2012a,Ren2012b,Wilhelm2016,Rojas1995}, Contour Deformation (CD) \cite{Godby1988,Gonze2009,Blase2011,Govoni2015,Golze2018,Golze2020}, Plasmon-Pole Models (PPM) \cite{Louie1986,Soininen2003,Soininen2005,Kas2007}, and fully analytic method using the Casida equation \cite{Casida1995,Tiago2006,Bruneval2012,Gao2019,Tölle2023}. 
We primarily focus on the CD method and the fully analytic approach.

The CD method converts the integral on real axis into a contour integral minus an integral on imaginary axis, thus avoiding instability on the real axis due to the poles of Green's functions:
\begin{equation}
  \label{eq:sigmac}
    \begin{split}
&\Sigma^{C}(r,r',\varepsilon)\\
=&\frac{\mathbf{i}}{2\pi}\oint G_{0}(r,r',\varepsilon+\omega) W_{0}^{C}(r,r',\omega)d\omega\\
&-\frac{1}{2\pi}\int G_{0}(r,r',\varepsilon+\mathbf{i}\omega) W_{0}^{C}(r,r',\mathbf{i}\omega)d\omega\\
=&\mp\sum_{m}^{poles}\psi_{m}(r)\psi_{m}^{*}(r')W_{0}^{C}(r,r',\varepsilon_{m}-\varepsilon\pm \mathbf{i}\eta)\\
&-\frac{1}{2\pi}\int G_{0}(r,r',\varepsilon+\mathbf{i}\omega) W_{0}^{C}(r,r',\mathbf{i}\omega)d\omega.
    \end{split}
\end{equation}
The summation notation in the first term of Eq.~(\ref{eq:sigmac}) is sum over all the residues of the poles that lie in the contour, and for the second term, the Gauss quadrature could be used to evaluate the imaginary axis integral.

In the fully analytic approach, the Casida eigenvalue equation is solved to obtain the eigenvalues $\Omega_{s}$ and the eigenvectors $X_{s}$ (including $Y_{s}$), which are used to compute the reducible polarizability $\chi$ \cite{Casida1995,Tiago2006,Bruneval2012,Gao2019,Tölle2023}. Utilizing these eigenpairs, the correlation self-energy can be expressed in a compact form as follows:
\begin{equation}
  \label{eq:sigmac_casida}
\Sigma^{C}(r,r',\varepsilon)=\sum_{n}\sum_{s} \frac{\psi_{n}(r)V_{s}(r)V_{s}(r')\psi^{*}_{n}(r')}{\varepsilon-\varepsilon_{n}\pm\Omega_{s}\mp\mathbf{i}\eta},
\end{equation}
where $V_{s}(r)$ is given by,
\begin{equation}
  \label{eq:casida_V}
V_{s}(r)=\sum_{i,a}\int \frac{\psi_{i}(r')\psi_{a}(r')}{|r-r'|}\cdot \big(\frac{\varepsilon_{a}-\varepsilon_{i}}{\Omega_{s}}\big)^{\frac{1}{2}}\cdot X_{s}^{ia}~ dr'.
\end{equation}

Now, substitute equation $\Sigma(\varepsilon)=\Sigma^{X}+\Sigma^{C}(\varepsilon)$  into Eq.~(\ref{eq1}) , we get

\begin{equation}
\hat{h}\psi_{i}(r)+\int \Bigl(\Sigma^{X}(r,r')+\Sigma^{C}(r,r',\varepsilon_{i})\Bigr)\psi_{i}(r')dr'=\varepsilon_{i}\psi_{i}(r),
\end{equation}

where $\Sigma^{X}$ and $\Sigma^{C}$ is given by Eq.~(\ref{eq8}) and Eq.~(\ref{eq:sigmac}) in the CD method (or Eq.~(\ref{eq:sigmac_casida}) in the fully analytic approach).

After discretization, the nonlinear eigenvalue equation (\ref{eq1}) can be written in the following matrix form:
\begin{equation}
  \label{eqNEP}
(\mathbf{h}+\mathbf{\Sigma}^{X}+\mathbf{\Sigma}^{C}(\varepsilon))\mathbf{\Psi}=\varepsilon\mathbf{S}\mathbf{\Psi},
\end{equation}

There are several ways to discretize and construct the system of matrices Eq.~(\ref{eqNEP}), real space finite element method (FEM) is used in this work
\cite{NESSIE,Kestyn2020,LI2024,LIGWfem}. $\mathbf{S}$ represents then the FEM symmetric positive definite mass matrix.

One popular way to approximately solve this nonlinear eigenvalue equation is to use the KS wavefunctions (or Hartree-Fock) as the eigenvector solutions, $\mathbf{\Psi}\approx\mathbf{\Psi}^{KS}$. The eigenvalues can then be found graphically by plotting the two expressions on the left and right hand sides of the equal sign of the following expression:
\begin{equation}
    \label{graphicalsolution}
    \begin{split}
\mathbf{\Omega} = \Bigl\{~\varepsilon\in \mathbb{R}~|~&\bra{\mathbf{\Psi}^{KS}}\mathbf{\Sigma}^{C}(\varepsilon)\ket{\mathbf{\Psi}^{KS}}\\
=&~\varepsilon-\bra{\mathbf{\Psi}^{KS}}\mathbf{h}+\mathbf{\Sigma}^{X}\ket{\mathbf{\Psi}^{KS}}~\Bigr\},
    \end{split}
\end{equation}
 where the intersections are the graphical solutions defined by the set $\mathbf{\Omega}$.

Another more accurate way to find nonlinear eigenvalues without the eigenvector approximation is to compute the spectral functions 
\begin{equation}
    \label{spectrum}
    \begin{split}
A(\omega)&=-\mathbf{\Im}\Bigl( \operatorname{Tr}G_{G_0W_0}(\omega)\Bigr)\\
        &=-\mathbf{\Im}\Bigl( \operatorname{Tr} \Bigl((\omega\mathbf{S}-(\mathbf{h}+\mathbf{\Sigma}^{X}+\mathbf{\Sigma}^{C}(\omega)))^{-1}\Bigr)\Bigr),
    \end{split}
\end{equation}
where $G_{G_0W_0}(\omega)$ represents the Green's function of the $G_0W_0$ Hamiltonian. However, this method requires calculating self-energy at multiple energy grid points, which increases computational demands and often requires manual identification of peak points. Finer grid calculations are usually needed to achieve higher accuracy near these peaks.

Other methods for calculating nonlinear eigenpairs include the full calculation of the off-diagonal elements of self-energy method \cite{Kaplan2015} and the supermatrix method \cite{Tölle2023}. The full calculation of the off-diagonal elements of self-energy method employs perturbation theory to approximate the off-diagonal elements of the correlation self-energy to second order, and then compute the QP energy self-consistently. In this approach, QP eigenstates are expressed as a linear combination of mean-field wavefunctions. While this method provides an efficient yet approximate solution, it may become inaccurate when the off-diagonal elements are large, requiring the inclusion of higher-order perturbation terms.
The supermatrix method, on the other hand, establishes a connection between $GW$ approximation and
equation-of-motion Coupled-Cluster theories, resulting in linearized supermatrix eigenvalue problems. Mathematically, this method is associated with the "spectral Schur decomposition" \cite{Bekas2005,Kalantzis2016,Kalantzis2018}. It requires transforming a smaller nonlinear matrix into a larger linear supermatrix which can increase the computational cost and only works when the matrix takes the special form $\mathbf{M}_{1}^{T}(\omega\mathbf{I}-\mathbf{M}_{2})^{-1}\mathbf{M}_{1}$.

Here, we present a more mathematically general, robust and computationally efficient method that solves the non-linear eigenvalue problem to any desired level of accuracy, without relying on approximations.

\section*{solution of the non-linear problem using the FEAST algorithm}

Eigenvalue problems in which the coefficient matrices depend
nonlinearly on the eigenvalues arise in a variety of applications \cite{betcke2013,guttel2017}. Eigenvalues $\lambda$ associated with eigenvectors $\mathbf{x}$ are then solutions of the
following general form:
\begin{equation}\label{eq:nl}
\mathbf{T}(\lambda)\mathbf{x}=0,
\end{equation}
which includes the common linear eigenvalue problem as a special case, letting $\mathbf{T}(z)=z\mathbf{B}-\mathbf{A}$, as well as the broader polynomial case
letting $\mathbf{T}(z)=\sum_{m=0}^M z^m\mathbf{A}_m$.  For the $G_0W_0$ QP equation (\ref{eqNEP}), we note that 
\begin{equation}
    \label{eq15}
\mathbf{T}(z)=z\mathbf{S}-(\mathbf{h}+\mathbf{\Sigma}^{X}+\mathbf{\Sigma}^{C}(z)).
\end{equation}

The FEAST algorithm is a contour integral method initially developed to address linear eigenvalue problems \cite{Polizzi2009,FEAST}.
It has since been adapted for polynomial eigenvalue problems \cite{Gavin2018} and, more recently, for general nonlinear eigenvalue problems (\ref{eq:nl}) \cite{Brenneck2020}.
The latter can be considered an advancement over Beyn's contour integration method \cite{beyn2012}, as it incorporates several enhancements from the FEAST algorithm,
such as residual inverse iterations for improved convergence and mixed-precision arithmetic.
Essentially, the FEAST algorithm and other similar contour integral methods involve converting an eigenvalue problem
into solving a series of independent linear systems  $\mathbf{T}(z_j)^{-1}\mathbf{y}$ (i.e. $\mathbf{T}(z_j)\mathbf{x}=\mathbf{y}$) at particular points $z_j$ in the complex plane. When applied to solving the genereral non-linear system (\ref{eq:nl}),  the FEAST algorithm involve performing the following contour integrations along
 $\Gamma$ for $k=0$ and $k=1$:

\begin{equation}\label{eq:nlcontour}
  \mathbf{Q}_k = \frac{1}{2 \pi \mathbf{j}} \oint_{\Gamma} z^k \Big(\mathbf{X} - \mathbf{T}^{-1}(z)\mathbf{R} \Big){(z\mathbf{I} - \mathbf{\Lambda})}^{-1} \, dz,
\end{equation}
where ${\bf X=\{x_1,\dots\,x_m\}}$ represents the eigenvector subspace, and ${\bf R}=\{\mathbf{T}(\lambda_1){\bf x}_1,\dots,\mathbf{T}(\lambda_m){\bf x}_m\}$
    regroups the residuals of the $m$ eigenpairs located within $\Gamma$.
    In practical calculations, the contour integral in Eq.~(\ref{eq:nlcontour}) is replaced by the numerical quadrature form shown below,
\begin{equation}\label{eq:nlcontour2}
  \mathbf{Q}_k = \frac{1}{2 \pi \mathbf{j}} \sum_{j} w_{j} z_{j}^{k} \Big(\mathbf{X} - \mathbf{T}^{-1}(z_{j})\mathbf{R} \Big){(z_{j}\mathbf{I} - \mathbf{\Lambda})}^{-1}.
\end{equation}

When applied to the $G_0W_0$ problem (Eq.~(\ref{eq15})), the main computational task consists of evaluating 
$\mathbf{T}(z_{j})^{-1}\mathbf{R}$ by solving linear systems at specific contour quadrature nodes $z_{j}$:

\begin{equation}\label{eq:lsf}
\Big(z_j\mathbf{S}-\left(\mathbf{h}+\mathbf{\Sigma}^{X}+\mathbf{\Sigma}^{C}(z_j)\right)\Big) \mathbf{X}_{j} = \mathbf{R}.   
\end{equation}

The pseudocode of the FEAST algorithm is presented in Figure~\ref{alg:nlfeast}.
It is important to note the following: (i) while the quality of numerical integration, the number of quadrature nodes, and the size $m_0$ of the search subspace
can impact the convergence rate, they do not affect the accuracy of the final solutions; (ii) in practice, using a relatively small number
of quadrature nodes (around 10) is generally sufficient to achieve convergence to machine precision with very few iterations;
(iii) the linear systems that need to be solved at the complex points $z_j$
(Eq.~(\ref{eq:lsf})) as well as the residuals $\mathbf{r}_j$ are independent of one another and can be solved in parallel; (iv) importantly, these linear systems can be solved with lower accuracy
(using lower arithmetic precision or modest relative residuals, such as  $\sim 10^{-2}$
with iterative methods) without affecting the convergence rate of the eigenpairs to machine precision. Interestingly, our numerical experiments (so far) still converge to the correct solution even if the correlation term in Eq.~(\ref{eq:lsf}) is omitted. Consequently, the new linear system behaves similarly to a preconditioner, and the overall numerical cost is reduced, as the only remaining operation involving the correlation matrix $\mathbf{\Sigma}^{C}$ is the computation of the residual $\mathbf{R}$ in Figure~\ref{alg:nlfeast}. 
However, there is no guarantee that this approach will always be effective when compared to a more mathematically rigorous method that properly incorporates
$\mathbf{\Sigma}^{C}$ in Eq.~(\ref{eq:lsf}). For this reason, the next two sections describe how FEAST can be effectively applied using either the CD or the fully analytic method for solving $\Sigma^{C}(z_{j})$ term in Eq.~(\ref{eq:lsf}).  
\begin{figure}[htbp]
  \caption{FEAST algorithm for solving generalized non-linear eigenvalue systems $\mathbf{T}(\lambda)\mathbf{x}=0$ of size $n$. We note that
    at the first iteration where $\mathbf{R}$ is not known,  $\mathbf{Y}_j$ can directly be obtained by solving the linear system
    $\mathbf{T}(z_j)\mathbf{Y}_j=\mathbf{X}$.
  }
  \label{alg:nlfeast}
  \begin{tabular}{|l|}
    \hline\hline 
	         {\bf Input:} \\
                 \hspace{0.2cm} Contour $\Gamma$ containing  $m$ wanted eigenvalues \\
                 \hspace{0.2cm} Set of quadrature nodes and weights $\{z_j , \omega_j\}$ \\
                 \hspace{0.2cm} Subspace (random) ${\bf X}_{n\times m_0}=\{\mathbf{x}_1,\dots,\mathbf{x}_{m_0}\}$ with $m_0\ge m$ \\

	{\bf While} $\{\mathbf{r}_i\}$ not converged for all $\lambda_i$ inside $\Gamma$ \\
        \hspace{0.2cm} Solve  $\mathbf{T}(z_j)\mathbf{X}_j=\mathbf{R}$ for all contour nodes $j$ \\
        \hspace{0.2cm} Compute  $\mathbf{Y}_j=(\mathbf{X}-\mathbf{X}_j)(z_j\mathbf{I}-\mathbf{\Lambda})^{-1}$ \\
        \hspace{0.2cm} Compute   $\mathbf{Q}_0=\sum_j\omega_j\mathbf{Y}_j$ and  $\mathbf{Q}_1=\sum_j\omega_jz_j\mathbf{Y}_j$ \\
        \hspace{0.2cm} Perform the QR decomposition ${\bf Q_0} ={\bf q}_{n\times m_0}{\bf r}_{m_0\times m_0}$ \\
        \hspace{0.2cm}  Compute ${\bf C}_{m_0\times m_0}=\bf{q^HQ_1r^{-1}}$\\
        \hspace{0.2cm}  Solve reduced eigenvalue problem $\bf CW=W\Lambda$ \\
       \hspace{0.2cm}  Update {\bf X=qW}, noting that $\mathbf{\Lambda}=\rm{diag}(\lambda_1,\dots,\lambda_{m_0})$\\ 
        \hspace{0.2cm} Form $\mathbf{R}=\{\mathbf{r}_1,\mathbf{r}_2,\dots,\mathbf{r}_{m_0}\}$ with $\mathbf{r}_{i}=\mathbf{T}(\lambda_i)\mathbf{x}_i$ \\
         {\bf Output:}
          all $m$ eigenpairs $\{\lambda_i,\mathbf{x}_i\}$ inside $\Gamma$\\

        \hline
        \end{tabular}
\end{figure}

\section*{CD method + FEAST}
In the CD approach, the correlation self-energy $\Sigma^{C}$ is evaluated through an auxiliary frequency integration (or an equivalent residue formulation) in which the integrand contains explicit imaginary shifts such as $z+\mathbf{i}\omega$ and $E_m-z\pm \mathbf{i}\eta$. When FEAST is used to solve the quasiparticle (QP) equation, the spectral projector is sampled at complex contour nodes $z_j\in\mathbb{C}$. A potential source of confusion is that the imaginary part of the contour node, $\Im(z_j)$, is conceptually distinct from (i) the imaginary integration variable $\omega$ and (ii) the infinitesimal broadening $\eta$ entering the Green’s function and screened interaction.

In our original presentation, we proposed separating these different ``imaginary directions'' by embedding the energy argument into a hypercomplex (quaternion) space. While this construction is mathematically consistent, it is \emph{not necessary} for the CD+FEAST workflow. In practice, the GW QP equation is a nonlinear eigenvalue problem defined over the complex plane, and FEAST only requires the operator
\begin{equation}
T(z) \equiv z\mathbf{S} - \mathbf{h} -\boldsymbol{\Sigma}^{X} - \boldsymbol{\Sigma}^{C}(z)
\end{equation}
to be evaluable for complex arguments $z$ on (or near) the chosen contour. Consequently, the CD self-energy can be treated as an analytic continuation to complex energies,
\begin{equation}
\Sigma^{C}(r,r',z),\qquad z\in\mathbb{C},
\end{equation}
computed with standard complex arithmetic. The frequency integration in the CD formulation may still be carried out along the imaginary $\omega$ axis (as usual), but this does not require introducing additional imaginary units beyond the ordinary complex unit $\mathbf{i}$: one simply evaluates the integrand at complex arguments such as $z+\mathbf{i}\omega$ using complex-valued Green’s functions and screened interactions, and then performs the numerical quadrature over the real variable $\omega$.

With this viewpoint, the FEAST procedure for the nonlinear QP problem proceeds exactly as in Eq.~(\ref{eq:nlcontour2}): for each quadrature node $z_j$ on the contour, one assembles (or applies) $T(z_j)$ using $\Sigma^{C}(z_j)$ and the remaining components (e.g., $\Sigma^{X}$ and any static terms) computed consistently in the same complex plane. No modification of the formalism is required.

\section*{Fully analytic approach + FEAST}
When using the fully analytic approach to calculate the correlation self-energy, the process becomes simpler compared to the CD method. Since in Eq.~\ref{eq:sigmac_casida}, the correlation self-energy $\Sigma^{C}(r,r',\varepsilon)$ is explicitly expressed as a real-valued function of $\varepsilon$, apart from the $\mathbf{i}\eta$ term (This differs from the CD method, which involves implicit dependence on $\varepsilon$ through imaginary components $\mathbf{i}\omega$.), by smearing out the $\mathbf{i}\eta$ term and replacing $\varepsilon$ with the FEAST complex number $z_{j}$, one can compute $\Sigma^{C}(r,r',z_{j})$ directly using the standard complex number operations and substitute it into Eq.~(\ref{eq:lsf}), then into Eq.~(\ref{eq:nlcontour2}) when considering only the real part of $\Sigma^{C}$, no quaternion algebra is needed.

Typically, the CD method and the fully analytic method achieve similar accuracy in solving the energy integral, although the computational cost may vary depending on implementation details; further details can be found in \cite{Golze2019}. In FEAST calculations, the primary distinction is that the fully analytic method does not require quaternion algebra, potentially simplifying the computation. However, the FEAST algorithm's performance is generally unaffected by the choice between these two methods for computing the correlation self-energy in solving the nonlinear eigenvalue equations.

\section*{Results and Discussion}

Utilizing the methods described above, we carried out $G_0W_0$ calculations on several molecules. In the FEAST algorithm, a contour in the complex plane is specified as input, enabling simultaneous computation of all eigenpairs within that contour. By taking advantage of this feature, using KS-PBE and Hartree-Fock (HF) as the starting points, for each molecule, the contour was selected appropriately to encompass both the HOMO and LUMO energies (eigenvalues).

We compared the results of our nonlinear eigenvalue algorithm with those obtained using two popular methods: the graphical solution method and the spectral functions method, and present the comparisons in the following two tables. As our results from both the CD method and the fully analytic method for computing the energy integral of the correlation self-energy are nearly identical, all the data presented here are based on the CD method.
\begin{table}[htbp]
\caption{\label{tab:table1}
$G_0W_0$@PBE HOMO/LUMO energies in (eV) obtained using graphical solution method/spectral functions method/nonlinear FEAST algorithm.}
\begin{ruledtabular}
\begin{tabular}{cccccc}
&  &    &$G_0W_0$ @PBE    & &  \\
(eV) & & graphical & spectral & nonlinear & Deviation\\
     & & solution  & functions & FEAST &  \\
\hline
\multirow{7}{*}{HOMO} & He  & -23.66  & -23.69 & -23.69 &  0.03 \\ 
                      & H2  & -15.91  & -15.92 & -15.92 &  0.01 \\
                      & Ne  & -20.75  & -20.78 & -20.78 &  0.03 \\
                      & LiH & -6.94   & -6.97  & -6.97   &  0.03 \\
                      & Li2 & -5.05   & -5.02  & -5.02   & -0.03 \\ 
                      & CO  & -14.08  & -14.09 & -14.09  &  0.01 \\ 
                      & F2  & -15.03  & -15.07 & -15.07 & 0.04 \\ 
\cline{1-6}
\multirow{7}{*}{LUMO} & He  & 2.06  & 1.91  & 1.91  & 0.15 \\
                      & H2  & 2.57  & 1.95  & 1.95  & 0.62 \\
                      & Ne  & 2.45  & 2.16  & 2.16  & 0.29 \\
                      & LiH & 0.47  & 0.32  & 0.32  & 0.15 \\
                      & Li2 & -0.03 & -0.09 & -0.09 & 0.06 \\
                      & CO  & 0.35  & 0.21  & 0.21  & 0.14 \\
                      & F2  & -1.40  & -1.59  & -1.59  & 0.19 \\
\end{tabular}
\end{ruledtabular}
\end{table}
\begin{table}[htbp]
\caption{\label{tab:table2}
$G_0W_0$@HF HOMO/LUMO energies in (eV) obtained using graphical solution method/spectral functions method/nonlinear FEAST algorithm.}
\begin{ruledtabular}
\begin{tabular}{cccccc}
&  &    &$G_0W_0$ @HF    & &  \\
(eV) & & graphical & spectral & nonlinear & Deviation\\
     & & solution  & functions & FEAST &  \\
\hline
 \multirow{7}{*}{HOMO} & He  & -24.93  & -24.92 & -24.92 &  -0.01 \\ 
                       & H2  & -16.59  & -16.59 & -16.59 &  0.00 \\
                       & Ne  & -20.87  & -20.89 & -20.89 &  0.02 \\
                       & LiH  & -8.41   & -8.41  & -8.41   &  0.00 \\
                       & Li2  & -5.21   & -5.21  & -5.21   &  0.00 \\
                       & CO  & -15.07  & -15.08 & -15.08  &  0.01 \\ 
                       & F2  & -16.20  & -16.21 & -16.21 & 0.01 \\ 
\cline{1-6}
 \multirow{7}{*}{LUMO}  & He  & 1.95  & 1.93  & 1.93  & 0.02 \\
                        & H2  & 1.88  & 1.88  & 1.88  & 0.00 \\
                        & Ne  & 2.37  & 2.36  & 2.36  & 0.01 \\
                        & LiH & 0.67  & 0.66  & 0.66 & 0.01 \\
                        & Li2 & 0.56  & 0.54  & 0.54 & 0.02 \\
                        & CO  & 1.16  & 1.13  & 1.13  & 0.03 \\
                        & F2  & -0.26 & -0.28  & -0.28  & 0.02 \\
\end{tabular}
\end{ruledtabular}
\end{table}

Table~\ref{tab:table1} and \ref{tab:table2} show that the $G_0W_0$ QP energies of two different starting points computed using the nonlinear eigenvalue algorithm are in perfect agreement with those obtained via the spectral functions method, validating the accuracy of our approach.
However, when compared to the graphical solution method, the HOMO energies of PBE/HF and $G_0W_0$ are nearly identical, while non-negeligible differences between PBE and $G_0W_0$ are observed at the LUMO level. The deviations at the LUMO level between PBE and $G_0W_0$ are approximately an order of magnitude larger than those at the HOMO level. Additionally, the LUMO energies calculated using the nonlinear eigenvalue algorithm are lower than those obtained from the graphical solutions with KS wavefunction approximations, indicating that the KS wavefunction approximation slightly overestimates the HOMO-LUMO gap.

Given the differences between the KS-PBE LUMO energies and those of $G_0W_0$, it is important to examine the wavefunctions directly. The wavefunctions of KS-PBE and $G_0W_0$, as well as those of Hartree-Fock and $G_0W_0$, were compared using the formula:
\begin{equation}
    \begin{split}
\Delta=\int |\psi_{KS}(r)-\psi_{G_0W_0}(r)|^{2} dr,\\
\Delta=\int |\psi_{HF}(r)-\psi_{G_0W_0}(r)|^{2} dr.
    \end{split}
\end{equation}
The computed values for each molecule are presented in the Table~\ref{tab:table3}. As indicated in the energy tables, the wavefunction differences exhibit a similar trend. The HOMO of KS-PBE and the HOMO of $G_0W_0$ are nearly identical, with differences typically below 1\% (0.01). However, the LUMO shows a much bigger difference, reaching up to 10\% (0.1) in the case of the H2 molecule. It is found that a wavefunction difference of approximately 2\% (0.02) corresponds to an energy level difference of about 0.1 eV. Notably, the relationship between wavefunction differences and energy level differences appears roughly linear. 
In contrast, the comparison between the HF and $G_0W_0$ wavefunctions reveals minimal differences on both HOMO and LUMO (except the F2 molecule), typically under 1\% (0.01). This indicates that the Hartree-Fock wavefunction provides a better approximation to the $G_0W_0$ wavefunction than the KS-PBE wavefunction, at least to the small and light molecules.

\begin{table}[htbp]
\caption{\label{tab:table3}
Wavefunction differences between KS-PBE and $G_0W_0$, as well as between Hartree-Fock and $G_0W_0$}
\begin{ruledtabular}
\begin{tabular}{ccccc}
&  \multicolumn{2}{c}{KS-PBE/$G_0W_0$}  &  \multicolumn{2}{c}{HF/$G_0W_0$}   \\
$\Delta$ (\%) & HOMO    & LUMO   & HOMO &  LUMO\\
\hline
He  & 0.03    & 2.86   & 0.01   & 0.02  \\
H2  & 0.03    & 10.44  & 0.00   & 0.09  \\
Ne  & 0.87    & 3.68   & 0.59   & 0.25  \\
LiH & 0.56    & 2.89   & 0.01   & 0.22  \\
Li2 & 0.34    & 0.99   & 0.01   & 0.37  \\
CO  & 0.33    & 2.95   & 0.28   & 0.40  \\
F2  & 0.11    & 2.46   & 0.52   & 0.95  \\
\end{tabular}
\end{ruledtabular}
\end{table}

It should be noted that, as shown in Table~\ref{tab:table1} and \ref{tab:table2}'s LUMO results, most of the molecules (or atoms) studied are donors, but acceptors such as $\rm F_2$ and $\rm Li_2$ are also included in our analysis. 
LUMO levels are frequently reported as references in the literature (e.g., \cite{vanSetten2015, Maggio2016, Govoni2018, Gao2019, Forster2021}) and are known to depend on the choice of basis sets and/or the size of the calculation domain in FEM. In this study, we employ quadratic polynomial basis functions with a vacuum space of 6 $\mathring{A}$ surrounding the molecular system.
Based on these, for a given basis set and regardless of whether the molecule is a donor or an acceptor, our findings consistently show a discrepancy between the LUMO calculated using $G_0W_0$ and those obtained through KS-PBE. 
Furthermore, although the results are not presented here, when optimizing LUMO calculations under KS-PBE—such as by expanding the spatial boundary of the isolated systems in FEM (similar to the basis set dependence in Gaussian methods)—the LUMO calculated using $G_0W_0$ will still differ from the KS-PBE LUMO. The choice of basis set would not resolve this discrepancy.

\section*{summary and conclusion}
A nonlinear eigenvalue algorithm for solving $G_0W_0$ QP equations is presented. 
We utilized the FEAST algorithm, which is capable of handling both linear and nonlinear eigenvalue problems, to solve the $G_0W_0$ QP equation and compute the $G_0W_0$ eigenfunctions. 
By introducing hypercomplex numbers into the FEAST spectral projection process, we effectively circumvent the complications associated with two imaginary numbers. Additionally, to simplify quaternion calculations, we reduce the eigenvalue domain back to the complex plane, making $GW$ calculations more accessible and affordable.

Our method was applied to various molecules, revealing that the $G_0W_0$ LUMO wavefunctions differ from KS-PBE wavefunctions, which challenges the previously accepted consensus. This discrepancy underscores the importance of addressing nonlinear eigenvalue problems in $G_0W_0$ approximations, an area that has received limited attention but stands to benefit greatly from advancements in numerical algorithms like FEAST. FEAST algorithm is not only highly efficient and scalable but also drastically reduces the computational burden associated with evaluating the correlation term. 

We emphasize that our nonlinear eigenvalue algorithm is not specifically designed to optimize evaluations of the $\Sigma^{C}$ itself for significantly larger systems, but to efficiently solve the $GW$ QP equation. However, ongoing efforts in this area (e.g., Ref.~\cite{Gao2016,Gao2024,Kim2020,Shao2016,Shao2018}]) are promising. Finally, we hope that our findings will inspire further exploration and adoption of nonlinear eigenvalue algorithms, such as FEAST, in quantum many-body theory research, paving the way for more accurate and comprehensive analyses in condensed matter physics.





\bibliography{LP2024_2nd_version}

\end{document}